\documentclass[aps,pra,twocolumn,floatfix]{revtex4-2}
\usepackage[utf8]{inputenc}
\usepackage{multirow}
\usepackage{graphicx}
\usepackage{siunitx}
\usepackage{hyperref}
\usepackage[english]{babel}
\usepackage{braket}
\usepackage{amsmath}
\usepackage{graphicx}
\usepackage{siunitx}
\usepackage{nicefrac}
\usepackage[rgb]{xcolor}

\begin{document}
\title{The role of Coulomb anti-blockade in the photoassociation of long-range Rydberg molecules}

\author{Michael Peper}
\thanks{Present address: Department of Electrical and Computer Engineering, Princeton University, Princeton, New Jersey 08544, USA}
\affiliation{Department of Physics and Geoscience, University of Leipzig, 04109 Leipzig, Germany}
\author{Martin Trautmann}
\affiliation{Department of Physics and Geoscience, University of Leipzig, 04109 Leipzig, Germany}
\author{Johannes Deiglmayr}
\email{johannes.deiglmayr@uni-leipzig.de}
\affiliation{Department of Physics and Geoscience, University of Leipzig, 04109 Leipzig, Germany}

\date{\today}

\begin{abstract}
We present a new mechanism contributing to the detection of photoassociated long-range Rydberg molecules via pulsed-field ionization: ionic products, created by the decay of a long-range Rydberg molecule, modify the excitation spectrum of surrounding ground-state atoms and facilitate the excitation of further atoms into Rydberg states by the photoassociation light. Such an ion-mediated excitation mechanism has been previously called \textit{Coulomb anti-blockade}. Pulsed-field ionisation typically doesn't discriminate between the ionization of a long-range Rydberg molecule and an isolated Rydberg atom, and thus the number of atomic ions detected by this mechanism is not proportional to the number of long-range Rydberg molecules present in the probe volume. By combining high-resolution UV and RF spectroscopy of a dense, ultracold gas of cesium atoms, theoretical modeling of the molecular level structures of long-range Rydberg molecules bound below $n\,^2$P$_{3/2}$ Rydberg states of cesium, and a rate model of the photoassociation and decay processes, we unambiguously identify the signatures of this detection mechanism in the photoassociation of long-range Rydberg molecules bound below atomic asymptotes with negative Stark shifts. 
\end{abstract}

\maketitle

\section{Introduction}

Long-range Rydberg molecules (LRMs) are bound states of a Rydberg atom and a ground-state atom where the bond is the result of elastic scattering of the Rydberg atom's electron from a ground-state atom within its orbit~\cite{greeneCreationPolarNonpolar2000}. Since their first observation by photoassociation spectroscopy in an ultracold gas of rubidium about a decade ago~\cite{bendkowskyObservationUltralongrangeRydberg2009}, numerous studies have focused on their properties, such as the presence of very large permanent electric dipole moments \cite{liHomonuclearMoleculePermanent2011,boothProductionTrilobiteRydberg2015} or their large spatial extent, which was exploited to probe pair-correlation functions~\cite{whalenProbingNonlocalSpatial2019,whalenHeteronuclearRydbergMolecules2020}, polaron dynamics~\cite{schmidtMesoscopicRydbergImpurity2016} in degenerate quantum gases and  ultracold chemical reactions in dense gases ~\cite{niederprumGiantCrossSection2015,schlagmullerUltracoldChemicalReactions2016}. A detailed understanding of the structure and dynamics of LRMs is a prerequisite for the realization of proposals to create exotic states of matter via the photoassociation of LRMs~\cite{peperFormationUltracoldIon2020,hummelUltracoldHeavyRydberg2020a}.

\begin{figure}[tb]
	\centering
    \includegraphics[width=0.96\linewidth]{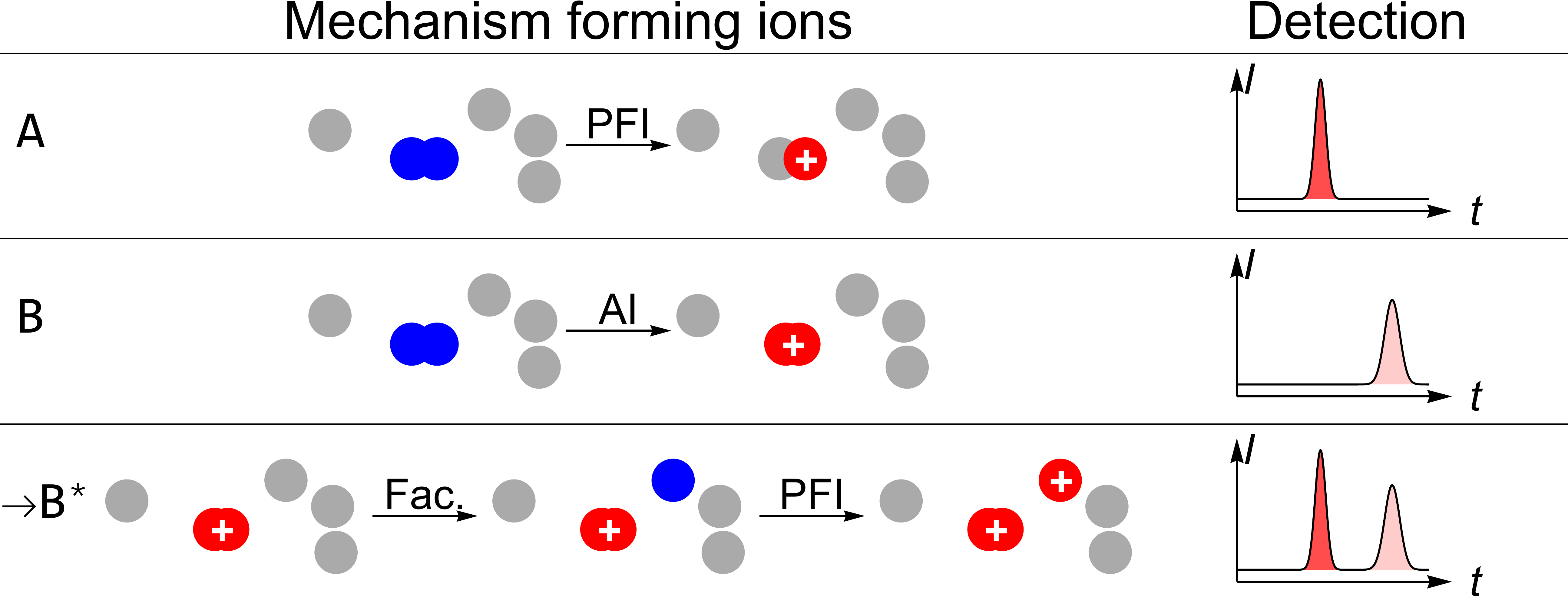}
	\caption{\label{fig:DetectionMechanism} Left column: schematic depiction of the three mechanisms by which long-range Rydberg molecules are detected in this work (filled circle: atom, overlapping circles: molecule, blue symbol: excited atom/molecule, red symbol: ionized atom/molecule, PFI: pulsed-field ionization, AI: molecular auto-ionization, Fac.: ion-facilitated excitation), for details see text. Right-hand column: illustration of the corresponding time-vs-charge traces recorded in the experiments ($I$: detector current, $t$: time-of-flight of the ion).}
\end{figure}

LRMs are formed by photoassociation in ultracold gases and detected by pulsed-field ionization (PFI) and the observation of the resulting ions by charged-particle detectors~\cite{marcassaChapterTwoInteractions2014,shafferUltracoldRydbergMolecules2018}. Figure~\ref{fig:DetectionMechanism} schematically depicts the different mechanisms through which (neutral) long-range Rydberg molecules (LRM) are detected using a charged-particle detector. In mechanism \textsf{A}, a LRM (blue dimer) is ionized by PFI, removing the Ryd\-berg electron. Because the Rydberg electron provides the binding of the LRM, its removal causes dissociation of the LRM into a neutral ground-state atom and a Cs$^+$ ion (red symbol). The resulting Cs$^+$ ion is accelerated by the ionization field towards the micro-channel-plate detector (MCP), causing a peak in the measured current after a distinct time of flight. In mechanism \textsf{B}, the LRM decays via vibrational auto-ionization to a stable Cs$_2^+$ molecular ion prior to PFI~\cite{niederprumGiantCrossSection2015,sassmannshausenLongrangeRydbergMolecules2016,schlagmullerUltracoldChemicalReactions2016}. The Cs$_2^+$ ion is also accelerated by the applied pulsed electric field towards the MCP. However, because the mass of the molecular ion is twice that of the atomic ion produced in mechanism \textsf{A}, the characteristic time of flight for mechanism \textsf{B} is approximately $\sqrt{2}$ times larger than for mechanism \textsf{A}. These two detection mechanisms are well known and have been used in previous works, \textit{e.g.}, to characterize the formation of heteronuclear LRMs~\cite{peperHeteronuclearLongRangeRydberg2021}. 

\begin{figure}[tb]
	\centering
    \includegraphics[width=0.84\linewidth]{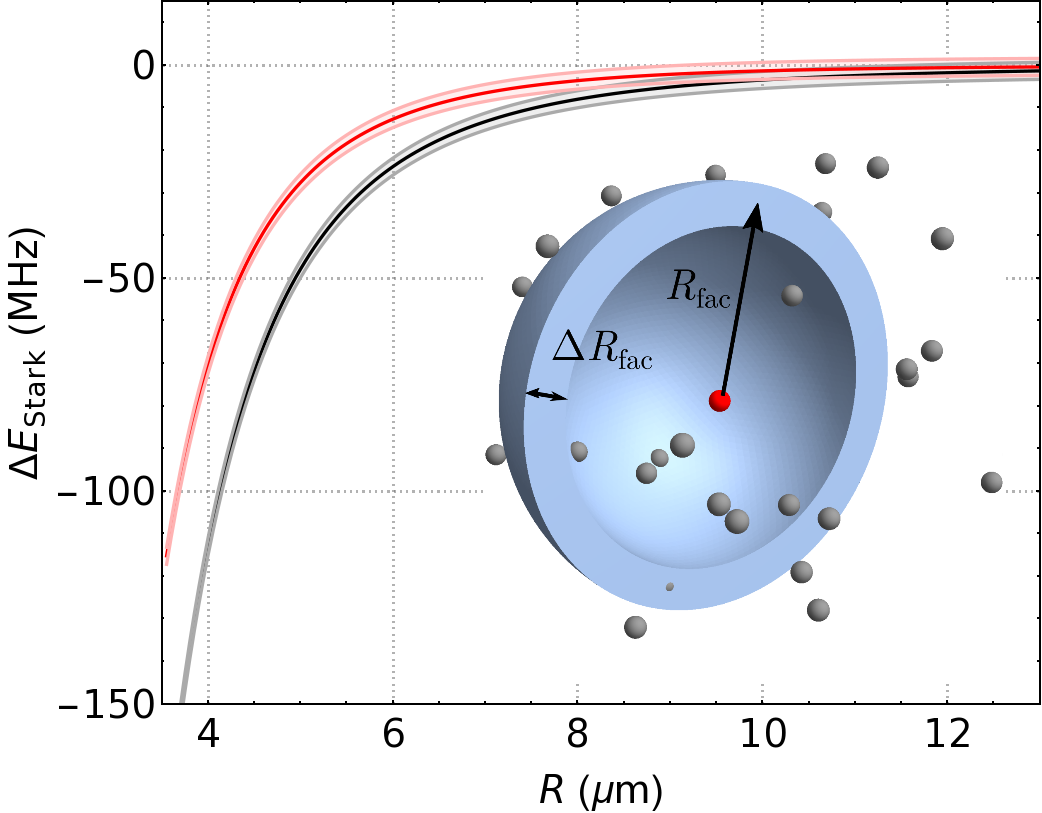}
	\caption{\label{fig:facilitationscheme} Stark shift of the $40\,^2$P$_{3/2}$ Rydberg state (black: $\Omega=1/2$, red: $\Omega=3/2$) in the spherical Coulomb potential of an ion at distance $R$, calculated following references \cite{duspayevLongrangeRydbergatomIon2021,deissLongRangeAtomIon2021}. The opaque bands illustrate the effect of an (exaggerated) excitation bandwidth of \SI{4}{\mega\hertz}. Inset: artistic impression of the facilitation shell with radius $R_\mathrm{fac}$ and thickness $\Delta R_\mathrm{fac}$ surrounding an ion (red ball) in a gas of ground-state atoms (gray balls).  }
\end{figure}

Here we report on a third detection mechanism \textsf{B$^*$}, ion-facilitated excitation of ground-state atoms into Rydberg states. This mechanism is initiated by the formation of a molecular ion through molecular autoionization (mechanism \textsf{B}) during the photoassociation laser pulse. The electric field of the ion influences  the excitation frequency of neighboring atoms through the Stark effect. Figure~\ref{fig:facilitationscheme} depicts the Stark shift of the $40\,^2$P$_{3/2}$ state of a cesium atom as a function of distance from a singly-charged ion. Because the frequency of the photoassociation laser is tuned below the atomic resonance by the binding energy of the addressed LRM, the transition from the ground state to the atomic Rydberg state becomes resonant at a certain internuclear distance $R_\mathrm{fac}$ between the ion and the ground-state atom as illustrated in Figure~\ref{fig:facilitationscheme}. 

The ion thus facilitates the resonant excitation of ground-state atoms into the $40\,^2$P$_{3/2}$ state, an enhancement effect for which previously the expression \textit{Coulomb anti-blockade} was coined~\cite{boundsCoulombAntiblockadeRydberg2019}. Subsequent PFI ionizes the excited atoms and, caused by the formation of a single LRM, an ion arriving at the time-of-flight of Cs$_2^+$, and one or more ions arriving at the time-of-flight of Cs$^+$ are observed. Noteworthy, this mechanism creates an ion signal at the same arrival time as PFI of LRMs, exclusively at molecular photoassociation resonances, but the initially formed LRM has already decayed before PFI. In contrast to a facilitated excitation of ground-state atoms through Rydberg-Rydberg interactions \cite{schemppFullCountingStatistics2014,malossiFullCountingStatistics2014}, this facilitation mechanism  does not cause an excitation avalanche because there remains only a single ion present in the system. The mechanism \textsf{B$^*$} is also closely related to the formation of the recently discovered class of ion--Rydberg-atom molecules, bound states of an ion and a Rydberg atom created by excitation of Rydberg-atoms in the vicinity of an ion with negligible kinetic energy \cite{duspayevLongrangeRydbergatomIon2021,deissLongRangeAtomIon2021,zuberObservationMolecularBond2022}. In our case, the molecular ion is formed by the autoionization process $\mathrm{Cs}_2\rightarrow \mathrm{Cs}_2^+ + e^-$ and has substantial kinetic energy: the binding energy of the formed Cs$_2^+$ ion can be up to $hc \times $\SI{5000}{\per\cm} \cite{jraijTheoreticalElectronicStructure2005,sassmannshausenLongrangeRydbergMolecules2016}, a fraction of \SI{4e-6}{} of which  (\textit{i.e.}, up to $h \times $\SI{600}{MHz}) is partitioned to the heavy ion. If the molecular ion is formed in low-lying vibrational levels, the initial kinetic energy of the formed ion-Rydberg--atom pair exceeds the potential energy at typical laser detunings and the pair formed by facilitated excitation is not bound.

\section{Experiment}

\begin{figure*}[tb]
	\centering
    \includegraphics[width=0.85\linewidth]{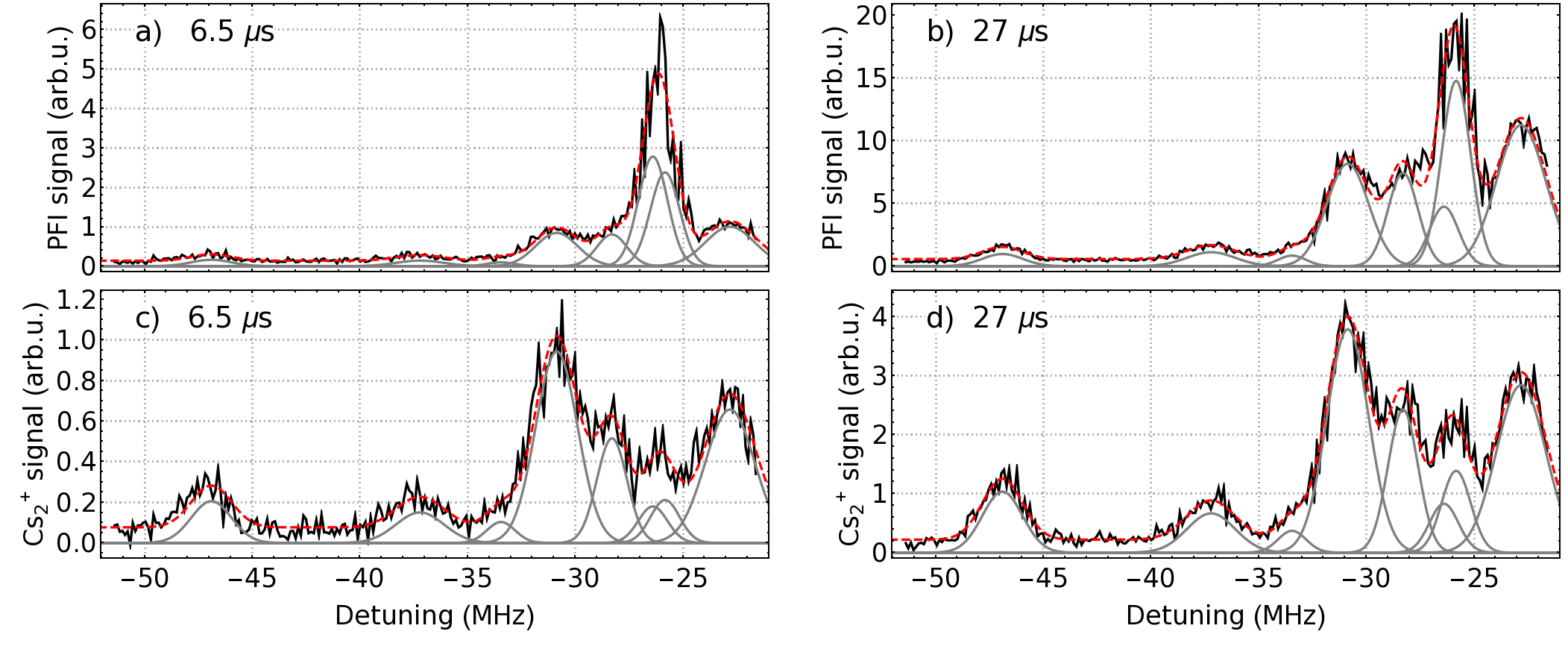}
	\caption{\label{fig:paspectrum40p} Ion signals detected after a photoassociation pulse with a length of \SI{6.5}{\micro\second} [a) and c)] and \SI{27}{\micro\second} [b) and d)], respectively, with frequencies below the atomic transition $40\,^2$P$_{3/2}\leftarrow6\,^2\mathrm{S}_{1/2}(F=3)$. a), b) (black line) signal detected in the time-of-flight window for Cs$^+$ ions created by pulsed-field ionization as function of the detuning of the UV-laser frequency from the atomic transition, (gray line) Gaussian line profiles fitted to the experimental spectrum, see Sect.~\ref{sec:padynamics} for details, (dashed-red line) sum of all Gaussian line profiles. c, d) (black line) signal detected in the time-of-flight window for Cs$_2^+$ ions created by autoionization, other lines as in a). Black and red-dashed curves have been vertically offset for clarity.}
\end{figure*}

To investigate the presence of ion-facilitated excitation in typical LRM photoassociation experiments, we study in detail the photoassociation dynamics close to the asymptote $40\,^2$P$_{3/2}$ in cesium. We note that the Rydberg-Rydberg interaction potentials between two cesium atoms in the $n\,^2$P$_{3/2}$ Rydberg state are repulsive for $n\leq 41$ \cite{mourachkoManyBodyEffectsFrozen1998}, resulting in a strong suppression of facilitation by Rydberg-Rydberg interactions for these states. Details of the experimental setup and procedures have been described in previous works~\cite{peperHeteronuclearLongRangeRydberg2021} and only the main aspects are summarized in the following. A cloud of \SI{2e7} cesium atoms with a density of \SI{1e11}{\per\cubic\cm} and a temperature of \SI{40}{\micro \kelvin} is prepared in the $6\,^2$S$_{1/2} (F=3)$ ground state by laser cooling in a magneto-optical trap, followed by magnetic compression and an optical molasses. A modified ring dye laser system (Coherent 899-21), pumped by a frequency-doubled continuous-wave Nd:YVO4 laser (Quantum finesse 532), and a frequency-doubling unit (Coherent MBD 200) are used to produce frequency-tunable light for the excitation into Rydberg states at wavelengths around \SI{319}{\nano\meter}. The frequency of the ring dye laser is stabilized to an atomic transition in potassium by a transfer cavity using a variable-offset-electronic-sideband locking technique~\cite{leopoldTunableLowdriftLaser2016}. Residual frequency drifts are on the order of \SI{100}{\kHz\per\day}. An AOM is used to create short pulses of variable intensity and length which are applied to the atomic sample. Excitation of Rydberg states is detected by PFI using ramped electric fields~\cite{sassmannshausenHighresolutionSpectroscopyRydberg2013}. For ground-state--Rydberg-state transitions, we observe typical line widths on the order of \SI{1.5}{MHz}~\cite{deiglmayrPrecisionMeasurementIonization2016}. 

Typical spectra are shown in Figure~\ref{fig:paspectrum40p} for photoassociation pulses of \SI{6.5}{\micro\second} [panels a)/c)] and \SI{27}{\micro\second} [panels b)/d)] length. In Figure~\ref{fig:paspectrum40p}\,a) and b), the time-of-flight signal was integrated in a time interval around the arrival time of Cs$^+$ ions and in Figure~\ref{fig:paspectrum40p}\,c) and d) around the arrival time of Cs$_2^+$ ions (see also Figure~\ref{fig:DetectionMechanism}). The spectra exhibit several resonances, some of which are only partially resolved. Photoassociation resonances are observed in all spectra at the same detunings but with clear differences in the intensity ratios. In the Cs$^+$-PFI spectra a) and b), the contrast between the strongest feature at a detuning of \SI{-26.4}{\MHz} and the other resonances is significantly reduced for the longer photoassociation pulse, hinting at the presence of a non-linear excitation or detection process. The Cs$_2^+$ spectra c) and d) have very similar intensity ratios, which, however, differ strongly from the ratios in the PFI spectra. Most notably, the resonance at \SI{-26.4}{\MHz} is much weaker than one would expect from the PFI spectrum.

\begin{figure}[tb]
	\centering
    \includegraphics[width=0.95\linewidth]{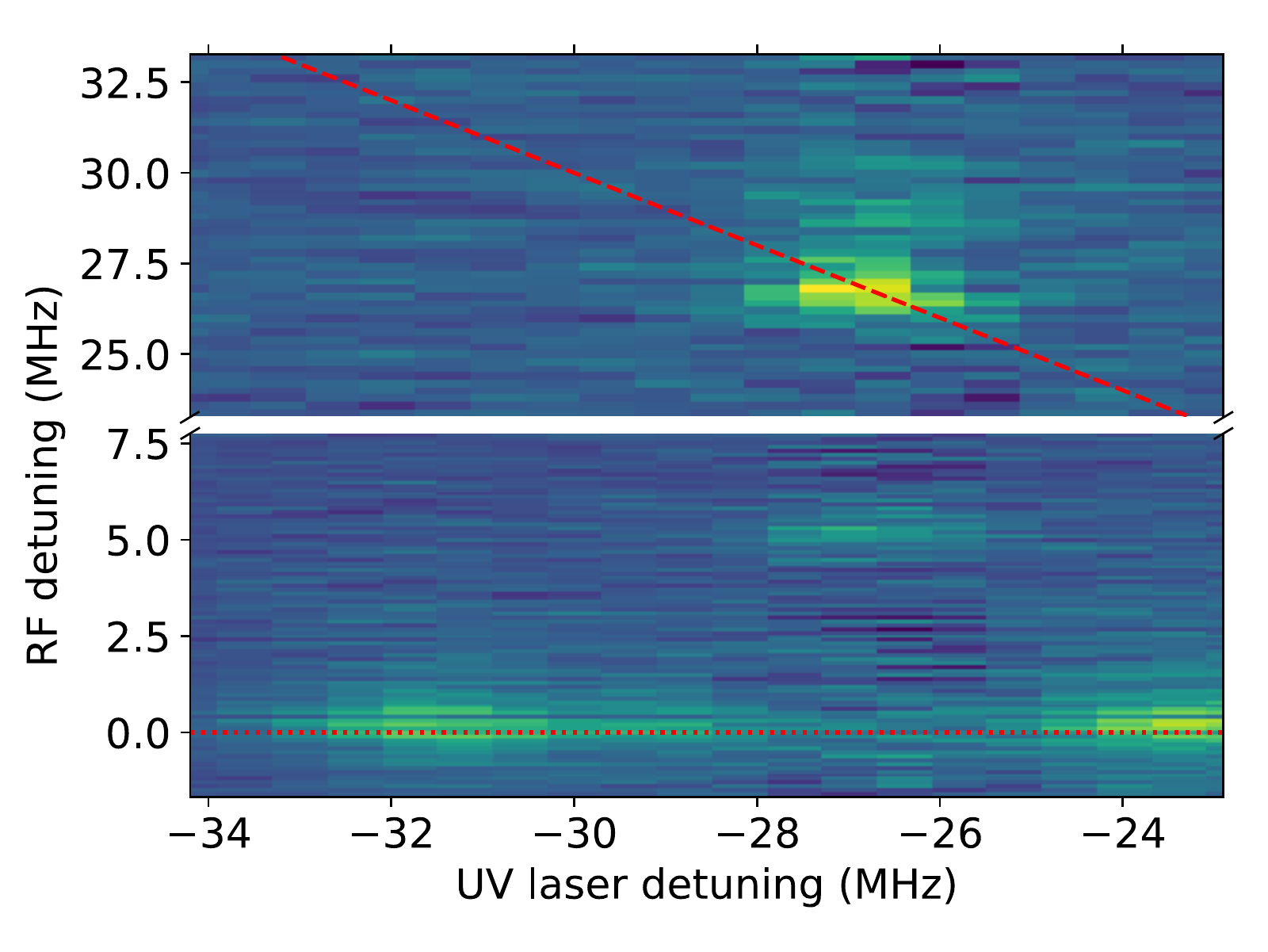}
	\caption{\label{fig:expRFsig} Fractional population in the $39\,^2$D$_{5/2}$ state after \SI{15}{\micro\second} photoassociation followed by \SI{2}{\micro\second} of RF transfer (blue: zero transfer, yellow: maximal transfer) as function of the detuning of the UV laser from the atomic transition $40\,^2$P$_{3/2}\leftarrow6\,^2\mathrm{S}_{1/2}(F=3)$ and the detuning of the RF field from the atomic transition $39\,^2$D$_{5/2}\leftarrow 40\,^2$P$_{3/2}$. Dashed line: transition frequencies for dissociation of LRMs. Dotted line: transition frequencies for transfer of isolated Rydberg atoms.}
\end{figure}

To investigate the contributions of mechanisms \textsf{A} and \textsf{B$^*$} to the detected PFI signal 
we perform RF spectroscopy of the sample after photoassociation~\cite{peperPhotodissociationLongrangeRydberg2020}. To this end, we apply a RF field with frequency close to the transition to the neighbouring $39\,^2$D$_{5/2}$ state after the photoassociation pulse. In this state, the Rydberg electron is less strongly bound and the atom or LRM ionizes at a smaller electric field compared to $40\,^2$P$_{3/2}$. Transfer between the atomic (molecular) Rydberg states is detected by applying a ramped electric-field-ionization pulse. The difference in binding energy causes a difference in the ionization time and consequently a shift in the arrival time of the resulting ion~\cite{sassmannshausenHighresolutionSpectroscopyRydberg2013} at the MCP. In Figure~\ref{fig:expRFsig}, the fractional population in the $39\,^2$D$_{5/2}$ state is color-coded and plotted as a function of the detuning of the UV laser from the atomic resonance (\textit{i.e.}, the binding energy of photoassociated LRMs) and the detuning of the RF frequency from the atomic transition. The data in the lower panel reveals the transfer from isolated atoms in the state $40\,^2$P$_{3/2}$ to the $39\,^2$D$_{5/2}$ Rydberg state, which is the signature of the ion-facilitation mechanism \textsf{B$^*$}. Because of the molecular binding energy, the RF transfer occurs for a LRM at different detunings than for an isolated atom. This is clearly visible in the upper panel of Figure~\ref{fig:expRFsig} for the photoassociation resonance at a UV laser detuning of \SI{-26.4}{\MHz}. At a positive RF detuning of \SI{26.4}{\MHz}, dissociation of LRMs into the continuum above the $39\,^2$D$_{5/2}$ asymptote is observed ~\cite{peperPhotodissociationLongrangeRydberg2020}. At lower frequencies, a bound-bound transition to a molecular state bound below the $39\,^2$D$_{5/2}$ asymptote results in a resonance at an RF detuning of about \SI{5}{\MHz}. This clearly indicates a dominant contribution from mechanism \textsf{A} to the PFI signal at this photoassociation resonance.

\section{Theoretical modelling of long-range Rydberg molecules}

We model the molecular level structure and experimental photoassociation spectra using the Hamiltonian of Eiles and Greene~\cite{eilesHamiltonianInclusionSpin2017}. Details of the numerical methods have been published previously~\cite{peperPhotodissociationLongrangeRydberg2020,peperHeteronuclearLongRangeRydberg2021} and only the main aspects are summarized here. We solve the Hamiltonian in the molecular frame including the hyperfine structure of the ground state and the fine structure of the Rydberg states in an atomic basis set consisting of the closest-lying low-$l$ Rydberg states and one degenerate manifold above and below the asymptote of interest on a grid of internuclear distances. The projection of the total angular momentum on the internuclear axis $\Omega$ remains a good quantum number and the calculations are performed separately for each value of $\Omega$. The electron-cesium scattering phases shifts are calculated following Khuskivadze \textit{et al.}~\cite{khuskivadzeAdiabaticEnergyLevels2002} where we have adjusted the parameters of the model-potential to reproduce experimental binding energies over a range of $n$ and $l$ asymptotes~\cite{peperAccuratePhaseShifts2022}. 

\begin{figure}[tb]
	\centering
    \includegraphics[width=0.95\linewidth]{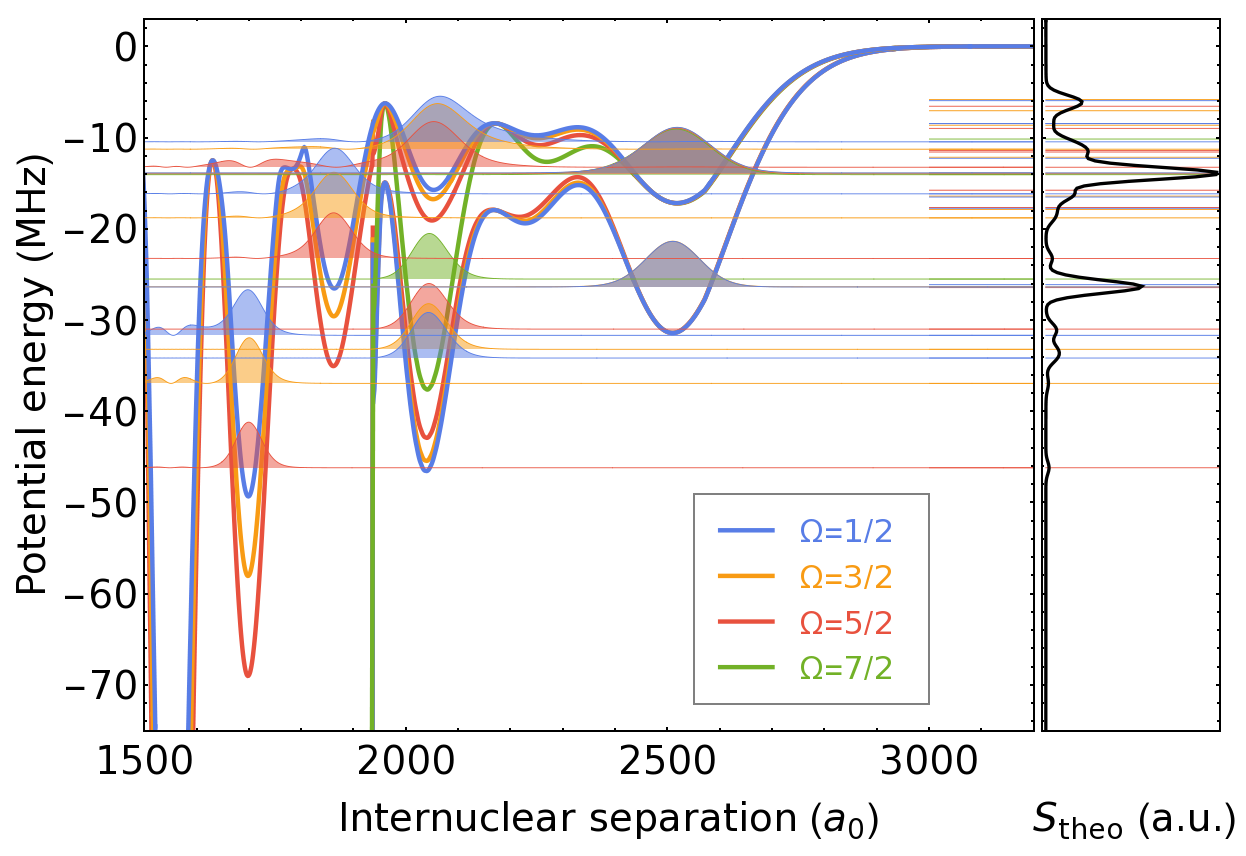}
	\caption{\label{fig:PEC40P} Left panel: (thick lines) calculated potential-energy curves (PECs) correlated to the asymptote $40\,^2$P$_{3/2}$-$6\,^2$S$_{1/2}(F=3)$. (Filled curves) vibrational wavefunctions of selected levels with normalized peak heights, scaled by a global factor to be displayed on the same scale as the PECs, and offset by the binding energy of the level. (Horizontal, thin lines): binding energies of all vibrational levels. Right-hand side panel: (black line) expected experimental spectrum, simulated as described in the text. (Horizontal, thin lines): binding energy of all vibrational levels.}
\end{figure}

We solve the nuclear Schr{\"o}dinger equation to obtain the energies, wavefunctions, and lifetimes of vibrational levels using the modified Milne-Phase amplitude method~\cite{sidkyPhaseamplitudeMethodCalculating1999}. The results for LRMs bound below the $40\,^2$P$_{3/2}$-$6\,^2$S$_{1/2}(F=3)$ asymptote are shown in Figure~\ref{fig:PEC40P} where all potential-energy curves and the wavefunctions of selected vibrational levels are drawn. The amplitudes of photoassociation resonances are approximated by the expression
\begin{equation}\label{eq:lineintegral}
    A_i =  \mathcal{C} \left(\int \Psi_i(R) R^2 \mathrm{d}R\right)^2 ,
\end{equation}
where $\Psi_i(R)$ is the normalized vibrational wavefunction of level $i$ and $\mathcal{C}$ is a global scaling constant absorbing units and experimental parameters such as the density of ground-state atoms and laser intensity. Equation~\ref{eq:lineintegral} assumes that the initial scattering wavefunction of the two colliding ground-state atoms is constant and coherent over the extent of the vibrational wavefunction of the photoassociated LRM. We find this assumption to reproduce experimental ratios of line strengths better than the assumption of an incoherent initial scattering state. An experimental photoassociation spectrum is then simulated as 
\begin{equation}
    S(\nu)=\sum_i^\mathrm{states}A_iV(\nu-\nu_i,\sigma,\Gamma_i)\,,
\end{equation}
where $V$ is a Pseudo-Voigt line profile with $\nu_i$ and $\Gamma_i$ being the binding energy and line width of molecular resonance $i$, respectively, and $\sigma$ being the effective experimental resolution.

Because of the high spectral density of vibrational levels and the sensitivity of their precise energies on the details of the model, we don't attempt to assign molecular quantum numbers to all observed photoassociation resonances. However, the comparison between the simulated spectrum in Figure~\ref{fig:PEC40P} and the experimental spectra in Figure~\ref{fig:paspectrum40p} yields important insights: \textit{i)} the resonance at \SI{-26.4}{\MHz} can be assigned to the formation of LRMs in the $v=0$ level of the outermost well of the more-strongly bound, triplet-scattering-dominated electronic states. The potential-energy curves for $\Omega = 1/2$, $3/2$, and $5/2$ are degenerated in this range of internuclear distances where $s$-wave scattering dominates, enhancing the predicted strength of the photoassociation resonance. The calculated lifetimes of these levels (with respect to tunneling to shorter internuclear distances and eventual vibrational auto-ionization) are much longer than the radiative lifetime of the atomic Rydberg state. This agrees very well with the conclusions drawn above from Figure~\ref{fig:expRFsig} that LRMs formed at a detuning of \SI{-26.4}{\MHz} have predominantly survived until detection by PFI (mechanism \textsf{A}) and explains the observation made in the discussion of Figure~\ref{fig:paspectrum40p} that much fewer Cs$_2^+$ ions are observed at this resonance compared to the other resonances. 

\textit{ii)} The other photoassociation resonances observed in the range of detunings depicted in Figure~\ref{fig:paspectrum40p} can be attributed to the formation of LRMs in the inner wells of electronic states. The contribution from $p$-wave scattering to the binding of these states causes spin-orbit interactions which lift the degeneracy in $\Omega$~\cite{deissObservationSpinorbitdependentElectron2020}, increasing significantly the spectral density of molecular states. The reduced extent of the tunneling barriers towards short internuclear distances in the inner wells results in short predicted lifetimes below few microseconds. This is consistent with the conclusion drawn from Figure~\ref{fig:expRFsig} that LRMs formed at the corresponding detunings predominantly autoionize before detection (mechanism \textsf{B}).

\textit{iii)} The simulated spectrum (Figure~\ref{fig:PEC40P}) reproduces qualitatively the experimental ratios between features \textit{i)} and \textit{ii)} in the PFI spectrum for a short photoassociation pulse [Figure~\ref{fig:paspectrum40p} a)]. For the longer photoassociation pulse, however, the PFI signals detected at short-lived molecular resonances [features \textit{ii)}] are disproportionately enhanced. This is further evidence for a contribution from mechanism \textsf{B$^*$} to the PFI signal detected at these resonances. 

\section{Dynamics of photoassociation and detection of LRMs}\label{sec:padynamics}

We systematically investigate the decay and facilitation dynamics by recording photoassociation spectra as shown in Figure~\ref{fig:paspectrum40p} for photoassociation pulses of different length and additionally insert a variable delay between photoassociation and detection by PFI. To extract the amplitudes of unresolved photoassociation resonances, we fit the spectra with an empirical line shape model. The model represents a photoassociation spectrum as a sum of Gaussian line-profiles, depicted as gray lines in Figure~\ref{fig:paspectrum40p}. Position and width of the individual resonances are determined from a fit to one selected spectrum where individual resonances are reasonably resolved. We then adjust this model to every recorded spectrum by only adjusting the individual amplitude, minimizing the squared sum of residuals. 

\begin{figure}[tb]
	\centering
    \includegraphics[width=0.96\linewidth]{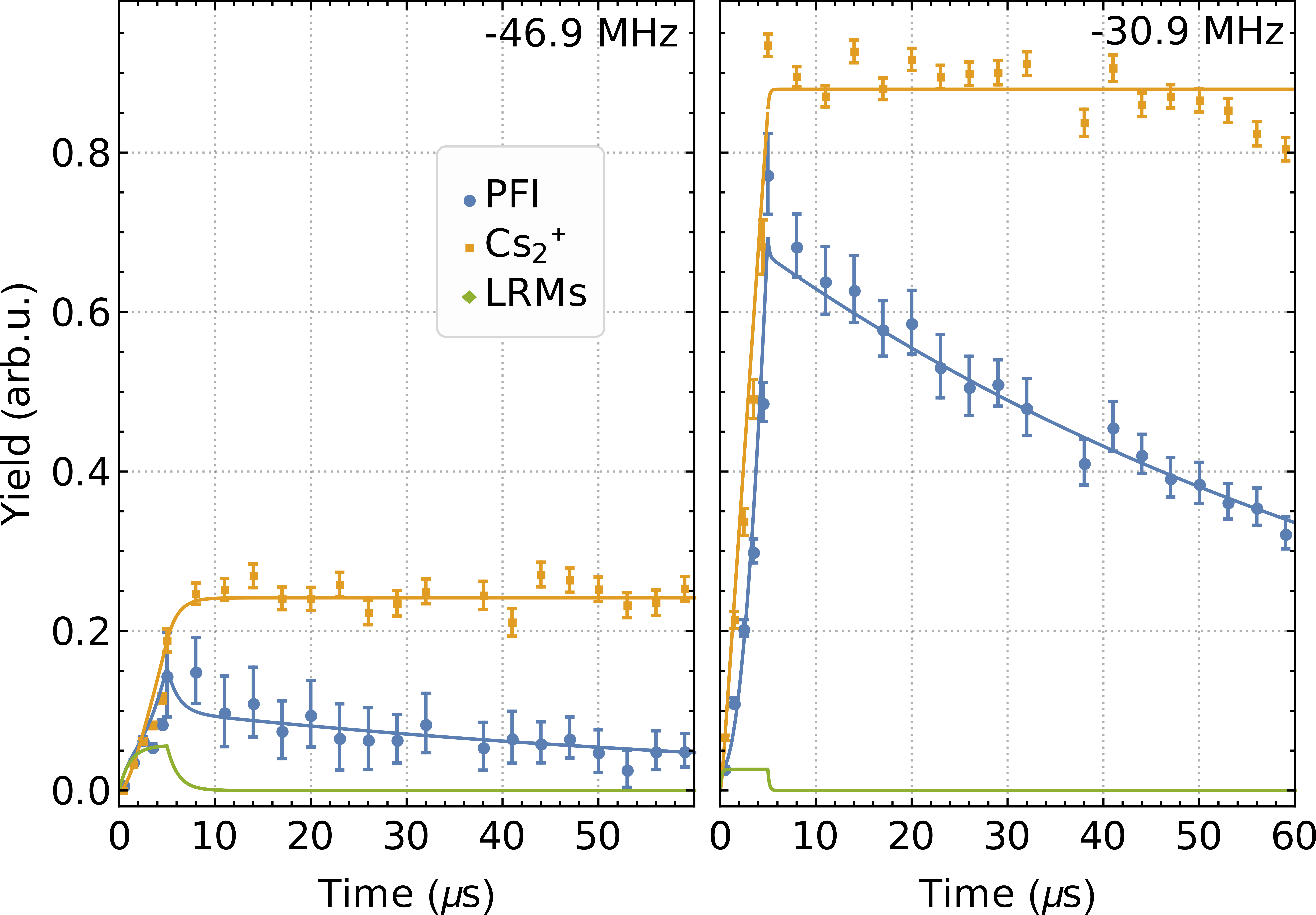}
	\caption{\label{fig:padynamics} Experimentally detected ion signals in the time-of-flight windows set for detection of ions produced by PFI (blue dots) and autoionization (orange dots) versus time of the PFI pulse (with respect to the beginning of the photoassociation pulse) for two selected resonances at the indicated detuning from the atomic transition $40\,^2$P$_{3/2}\leftarrow6\,^2\mathrm{S}_{1/2}(F=3)$. The data is compiled from two experimental sequences: in sequence 1 the length of the photoassociation pulse was varied up to \SI{5}{\micro\second}, followed directly by the PFI pulse, while in sequence 2 the length of the photoassociation pulse was kept fixed at \SI{5}{\micro\second} and an additional delay was inserted before PFI. Solid lines are results of weighted fits of the rate model \eqref{eq:ratemodel} to the combined data set: (blue) $N_\mathrm{LRM}(t) + N_\mathrm{Fac}(t)$, (orange) $N_\mathrm{Cs_2^+}(t)$, (green) only $N_\mathrm{LRM}(t)$.}
\end{figure}

Exemplary results of this procedure are depicted in Figure~\ref{fig:padynamics} for two resonances. We model the time-dependent signals for ions produced by PFI and Cs$_2^+$ ions with the following system of rate equations:
\begin{equation}\label{eq:ratemodel}
\begin{aligned}
\dot{N}_\mathrm{LRM}(t)&= k_\mathrm{PA} I(t) - k_\mathrm{AI} N_\mathrm{LRM}(t) - k_\mathrm{rad} N_\mathrm{LRM}(t) \\
\dot{N}_\mathrm{Cs_2^+}(t)&= k_\mathrm{AI} N_\mathrm{LRM}(t) \\
\dot{N}_\mathrm{Fac}(t)&= k_\mathrm{Fac} N_\mathrm{Cs_2^+}(t) -k_\mathrm{rad} N_\mathrm{Fac}(t) ,
\end{aligned}
\end{equation}
where $N_\mathrm{LRM}(t)$, $N_\mathrm{Cs_2^+}(t)$, and $N_\mathrm{Fac}(t)$ are the number of LRMs, Cs$_2^+$ ions, and facilitated atoms present at time $t$, respectively. $k_\mathrm{PA}$, $k_\mathrm{AI}$, $k_\mathrm{rad}$, and $k_\mathrm{Fac}$ are the rates for photoassociation, autoionization, radiative decay, and ion facilitation, respectively. $I(t)$ is a step function used to implement the length of the photoassociation pulse with respect to the total sequence described in the caption of Figure~\ref{fig:padynamics}. 

The model succeeds in describing the dynamics at all short-lived resonances, except for the spectral feature at \SI{-26.4}{\MHz} where several resonances with different formation and decay rates overlap. We thus focus in the following on the facilitation dynamics for short-lived LRMs. The model allows us to extract the number of remaining LRMs present in the system at every point in time, as depicted by the green model curves in Figure~\ref{fig:padynamics}. It becomes apparent that, at the end of the PA pulse, for nearly all of the observed molecular resonances only a small fraction of the detected PFI signals can be attributed to LRMs, while the major part of the signal originates from PFI of isolated Rydberg atoms in the state $40\,^2$P$_{3/2}$, excited via ion facilitation. This is in excellent agreement with the interpretation of the data from RF spectroscopy given above, and is further evidence for the contribution of mechanism \textsf{B$^*$} to the detection of LRMs by PFI.

The dependence of the facilitation rate $k_\mathrm{Fac}$ on the principal quantum number $n$ and the binding energy of the LRM, \textit{i.e.}, on the detuning of the UV laser from the atomic transition can be estimated from the Stark shift of the atomic transition generated by the electric field of the ion. For small Stark shifts and a homogeneous gas of ground-state atoms, the facilitation rate is proportional to the volume of a facilitation shell in which the photoassociation laser is in resonance with the Stark-shifted atomic transition. As shown in Figure~\ref{fig:facilitationscheme}, the radius of a facilitation shell depends on the laser detuning because of the inhomogeneous field created by the ion. We note that higher-order multipole moments of the ion's charge distribution contribute to the Stark shift and that the distance dependence of the Stark shift thus deviates from the $1/R^4$-scaling one would expect for a homogeneous field of strength $F=1/R^4$ (in atomic units) at the position of a ground-state atom~\cite{duspayevLongrangeRydbergatomIon2021,deissLongRangeAtomIon2021}. To gain fundamental insight and develop a simple model, however, we nevertheless approximate the Stark shift by the expression $\left(\nicefrac{1}{2}\right) \alpha_{nlj\Omega} R^{-4}$, where $\alpha_{nlj\Omega}$ is a Rydberg-state-dependent, effective dipole polarizability. For a laser frequency red detuned from the atomic asymptote with $\delta=\Delta E / h$ one therefore finds the mean radius of the facilitation shell $R_\mathrm{Fac}=\left(\frac{\alpha_{nlj\Omega}}{2\Delta E}\right)^{1/4}$. The width $\delta R$ of the facilitation shell is determined by the excitation bandwidth $\delta E$ as $\delta R = |\frac{\partial R_\mathrm{Fac}}{\partial \Delta E}| \delta E$ (see Figure~\ref{fig:facilitationscheme}). The volume of the facilitation shell, and thus the facilitation rate, then scales as 
\begin{equation}\label{eq:detdependence}
k_\mathrm{Fac}\propto V_\mathrm{Fac}\propto \frac{\alpha_{nlj\Omega}^{3/4}}{{\Delta E}^{7/4}} \hspace{0.3cm}. 
\end{equation}
Because of the general $n$-scaling of the polarizability $\alpha \propto n^7$ and the  binding energy of LRMs $\Delta E \propto n^{-6}$, this results in an approximate $n$-scaling of the facilitation volume as $V_\mathrm{fac}\propto n^{63/4} \sim n^{16}$. For the parameters of the measurement depicted in Figure~\ref{fig:padynamics} and a ground-state atom density of $n_\mathrm{gs}=\SI{1e11}{\per\cubic\centi\meter}$, at every instant of time a few ground-state atoms reside within the facilitation volume, justifying the use of the rate-equation model \eqref{eq:ratemodel}. Based on the strong scaling of the facilitation volume, and accordingly the effective facilitation rate, with the principal quantum number $n$, we expect that mechanism \textsf{B$^*$} is of general importance in the photoassociation of LRMs at large values of $n$ whenever LRMs decay to ionic products in the presence of the photoassociation light and the correlated atomic asymptote experiences a Stark shift to lower frequencies (\textit{e.g.}, the $n\,^2$P and $n\,^2$S low-$l$ Rydberg series of the alkali atoms).

\begin{figure}[tb]
	\centering
    \includegraphics[width=0.85\linewidth]{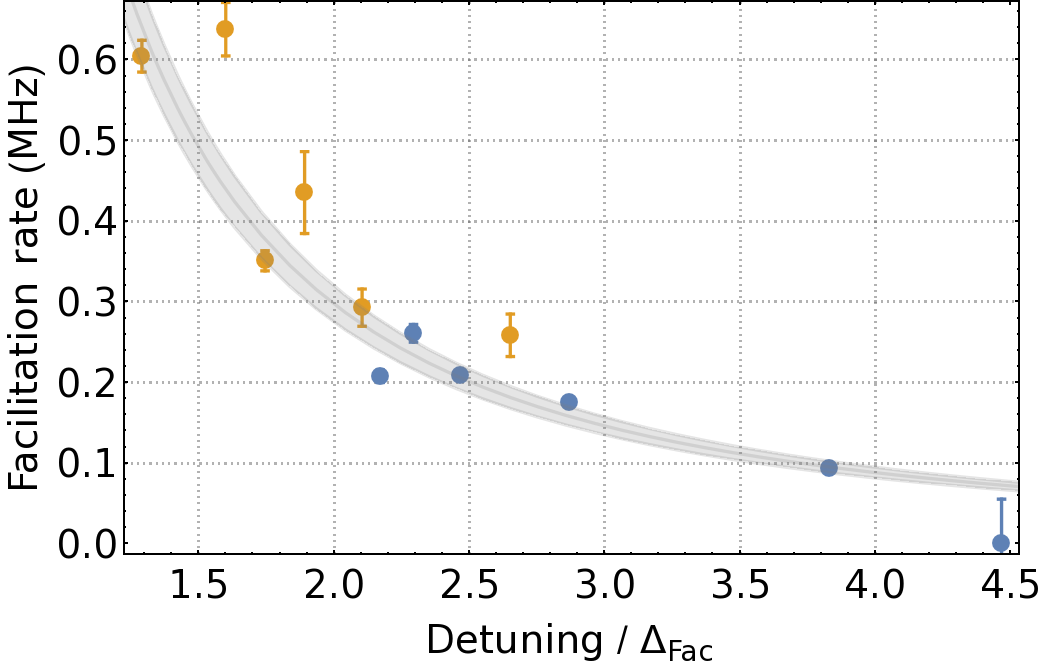}
	\caption{\label{fig:one} Extracted facilitation rates for photoassociation resonances below $31\,^2$P$_{3/2}$ (blue) and $40\,^2$P$_{3/2}$ (orange) as function of the effective detuning from the atomic resonance (see text for details). The gray line is a fit of the analytical model described in the text, the gray band indicates the \SI{95}{\percent} confidence interval.}
\end{figure}

In Figure~\ref{fig:one}, the facilitation rates $k_\mathrm{Fac}$, extracted from data sets similar to the ones shown in Figure~\ref{fig:padynamics} for all resonances in Figure~\ref{fig:paspectrum40p} and the analysis of a similar data set from measurements at $31\,^2$P$_{3/2}$, are shown as a function of the detuning of the photoassociation laser from the atomic transition. To eliminate differences in polarizability and experimental conditions between the two measurements, the facilitation rates of $31\,^2$P$_{3/2}$ and $40\,^2$P$_{3/2}$ are plotted against a normalized detuning $\delta / \Delta_\mathrm{Fac}$. $\Delta_\mathrm{Fac}$ was determined separately for the two values of $n$ by a weighted fit of Equation~\eqref{eq:detdependence} to the data and choosing $\Delta_\mathrm{Fac}$ such that at a detuning $\delta = \Delta_\mathrm{Fac}$ the facilitation rate would be \SI{1}{\MHz}. The gray curve in Figure~\ref{fig:one} is an independent fit of the power law  $k_\mathrm{Fac}\propto\delta^{-7/4}$ to the combined data sets, which captures the observed dependence well.

\section{Conclusions}

In this article, we demonstrated that vibrational autoionization of LRMs, followed by ion-facilitated excitation of ground-state atoms in the vicinity of the molecular ion (mechanism \textsf{B$^*$}), can enhance the observed ion yield at a photoassociation resonance: \textit{i)} RF spectroscopy unambiguously revealed the presence of isolated Rydberg atoms after photoassociation of LRMs with high autoionization rates. \textit{ii)} A detailed theoretical model of binding energies and line widths of photoassociation resonances consistently explained the observed autoionization dynamics. \textit{iii)} The observed dynamics of the photoassociation, decay, and detection process of LRMs was reproduced by a rate-equation model for the three detection processes \textsf{A}, \textsf{B}, and \textsf{B$^*$}. The extracted rates for ion-facilitated excitation of ground state atoms were found to be consistent with the scaling expected from simple arguments based on the Stark-shift of Rydberg states and the pair-distance distribution in a thermal gas.

While we have focused our study on long-range Rydberg molecules bound below $n\,^2$P$_{3/2}$ Rydberg states of cesium ($n=31,40$), we expect our results to be applicable to many other systems. The requirements for the observation of mechanism \textsf{B$^*$} are: a) photoassociation below a transition to an atomic Rydberg state with a negative Stark-shift, which includes the $n\,^2$P and $n\,^2$S low-$l$ Rydberg series of the alkali atoms, and b) the decay of the formed LRM during the photoassociation pulse, yielding an ionic product, which was observed in many previous experiments~\cite{sassmannshausenExperimentalCharacterizationSinglet2015,niederprumGiantCrossSection2015,schlagmullerUltracoldChemicalReactions2016,peperHeteronuclearLongRangeRydberg2021}. We note that our results might be of importance for the interpretation of previous experimental observations, such as surprisingly large probabilities for the formation of LRMs in thermal gases, where the ion-facilitation mechanism \textsf{B$^*$} might have caused an enhancement of the number of apparently detected LRMs~\cite{maclennanDeeplyBound24D2019}.

\begin{acknowledgments}
This work was supported by the Deutsche Forschungsgemeinschaft through SPP 1929 (GiRyd) under Project No. 428456632.
\end{acknowledgments}

\end{document}